\documentclass{book}    
\usepackage{piers}  
\usepackage[]{graphicx}
\graphicspath{{Figures/}}
\usepackage{subfigure}
\usepackage{amsmath}
\usepackage{amssymb}
\usepackage{epsfig}
\usepackage{cite}
\usepackage{color}
\usepackage{balance}
\usepackage{pgfplots}
\usepackage{hyperref}

\begin{document}

\title{Study of Clear Sky Models for Singapore}
\maketitle

\begin{authors}

{\bf Soumyabrata Dev}$^{1}$, {\bf Shilpa Manandhar}$^{2}$ $^{*}$, {\bf Yee Hui Lee}$^{3}$, {\bf and Stefan Winkler}$^{4}$ $^{**}$\\
\medskip

$^{1}$Nanyang Technological University Singapore, Singapore 639798, email: \texttt{soumyabr001@e.ntu.edu.sg}\\

$^{2}$Nanyang Technological University Singapore, Singapore 639798, email: \texttt{shilpa005@e.ntu.edu.sg}\\

$^{3}$ Nanyang Technological University Singapore, Singapore 639798, email: \texttt{EYHLee@ntu.edu.sg}\\

$^{4}$Advanced Digital Sciences Center (ADSC), Singapore 138632, email: \texttt{Stefan.Winkler@adsc.com.sg}

$^{*}$ Presenting author\\
$^{**}$ Corresponding author

\end{authors}

\begin{paper}

\begin{piersabstract}
The estimation of total solar irradiance falling on the earth's surface is important in the field of solar energy generation and forecasting. Several clear-sky solar radiation models have been developed over the last few decades. Most of these models are based on empirical distribution of various geographical parameters; while a few models consider various atmospheric effects in the solar energy estimation. 
In this paper, we perform a comparative analysis of several popular clear-sky models, in the tropical region of Singapore. This is important in countries like Singapore, where we are primarily focused on reliable and efficient solar energy generation. We analyze and compare three popular clear-sky models that are widely used in the literature. We validate our solar estimation results using actual solar irradiance measurements obtained from collocated weather stations. We finally conclude the most reliable clear sky model for Singapore, based on all clear sky days in a year. 
\end{piersabstract}

\psection{Introduction}
The total solar irradiance falling on the earth's atmosphere, in the event of the clear day without any clouds, is referred as the clear-sky solar radiation. It is important, particularly in tropical countries like Singapore, that experiences most of the direct solar irradiance~\cite{IGARSS_solar}, and is useful for solar energy generation. Several theoretical models have been developed that estimates the clear sky global solar irradiance (GHI). In this paper, we discuss three popular clear-sky models: Bird model~\cite{Bird81}, Yang model~\cite{dazhi2012estimation} and CAMS McClear model~\cite{McClear13}. 

\psection{Clear Sky Models}
Bird et al.\ uses results from radiative transfer codes, to estimate the total irradiance, by keeping the various atmospheric factors as constant throughout the year. It explains the various scattering of the solar radiation in the earth's atmosphere, and calculates the diffused- and direct- solar radiation at the earth's surface. The various atmospheric scattering from aerosols, ozone and water vapor are considered in its estimation of solar radiation. 

Recently, Yang et al.\ proposed an empirical regression model, that is specifically designed for Singapore. This model is created by considering three distinct weather station data positioned at various locations in Singapore. It modifies the clear-sky model in \cite{Robledo2000}, and proposes a Singapore-specific solar radiation model based on two user inputs -- zenith angle of the sun at the particular location on earth's surface and day number of the year.

The total clear-sky solar irradiance $G_c$ in W/$\mbox{m}^2$ is defined as: 
\begin{align}
\label{eq:GHI-model}
G_c = 0.8277E_{0}I_{sc}(\cos\theta_z)^{1.3644}e^{-0.0013\times(90-\theta_z)},
\end{align}

In Eq.~\ref{eq:GHI-model}, $E_{0}$ is the eccentricity correction factor of earth, $I_{sc}$ is a solar irradiance constant ($1366.1W/m^2$), $\theta_z$ is the solar zenith angle (in degrees). We compute the correction factor $E_{0}$ as follows:

\begin{equation*}
\begin{aligned}
\label{eq:E0value}
E_0 = 1.00011 + 0.034221\cos(\Gamma) + 0.001280\sin(\Gamma) + \\0.000719\cos(2\Gamma) + 0.000077\sin(2\Gamma).
\end{aligned}
\end{equation*}

$\Gamma = 2\pi(d_n-1)/365$ is the day angle (in radians), where $d_n$ is the day number in a year.

In addition to these empirical models, the atmosphere service of Copernicus (CAMS) proposed a Copernicus McClear clear-sky solar radiation model. It provides a fully physical model, by attempting to replace earlier empirical models. It uses the recent measurement results of various aeorosol properties and water vapor content, to provide reliable estimates of solar irradiance. The clear-sky solar radiation data are available for download from its site~\footnote{\url{http://www.soda-pro.com/web-services/radiation/cams-mcclear}.}, in temporal resolution of $1$ minute, $15$ minute, hour, day, and month.

\psection{Experiments and Results}
In this paper, we study these models~\footnote{The source codes of these models and the associated results are available online at \url{https://github.com/Soumyabrata/clear-sky-models}.} for a specific location in Singapore. In our experiments, we manually select all the clear-sky days in Singapore. However, we need to note that the existence of \emph{completely} clear-sky day is rare in a tropical country like Singapore. 

\psubsection{Data Collection}
The data is collected on a particular rooftop ($1.34^{\circ}$N, $103.68^{\circ}$E) of Nanyang Technological University Singapore. The total solar radiation is measured in resolution of $1$ minute, by Davis Instruments $7440$ Weather Vantage Pro II. In addition to solar radiation, it also provides us other useful meteorological data, such as temperature, humidity, precipitation and wind speed. We have also designed a ground-based sky camera called WAHRSIS (Wide Angled High Resolution Sky Imaging System)~\cite{WAHRSIS,IGARSS2015} that captures images of the sky scene at regular intervals of time. These sky cameras provide us more useful insights, while manually selecting the clear-sky days in a year. 

\psubsection{Results}
Figure~\ref{fig:csmodels} shows the various models for a sample clear sky day of 7-April-2016, along with the actual solar irradiance measurements, recorded by the weather station at this location. We observe that, for this sample day, both Bird model and Yang model over-estimates the actual solar irradiance values. However, McClear model can better estimate the actual solar irradiance measurements.

\begin{figure}[htb]
\centering
\includegraphics[height=0.4\textwidth]{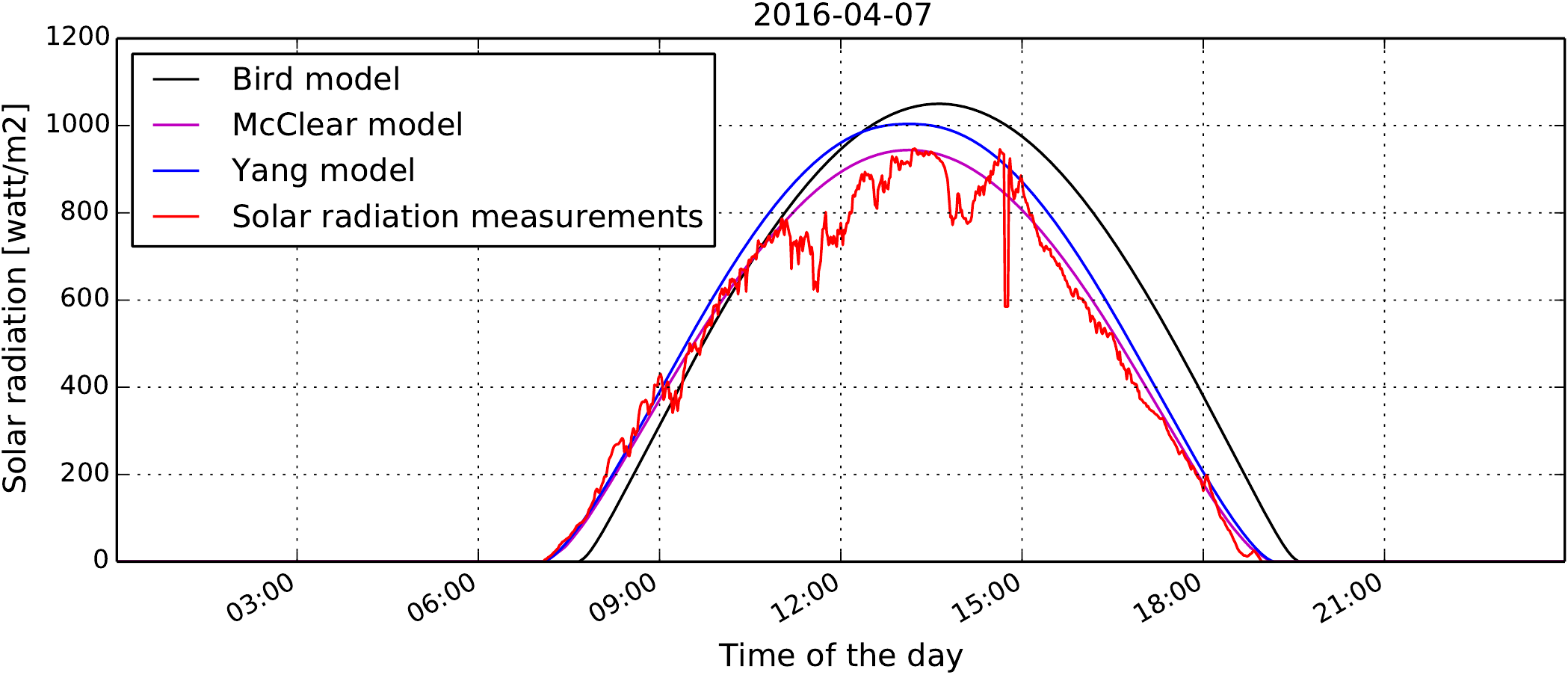}
\caption{Comparison of various clear sky models for a sample day in Singapore.}
\label{fig:csmodels}
\end{figure}

In order to provide an objective evaluation of the various models, we compute the Pearson correlation co-efficient and Root Mean Square Error (RMSE) between the actual- and estimated- solar irradiance values. Table~\ref{tab:oneday-results} shows the results for 7-April-2016. The McClear model has the least RMSE value, and highest correlation value amongst all the models. 

\begin{table}[htb]
\centering
\begin{tabular}{l|c|c}
\hline
              & Correlation & RMSE (W/$\mbox{m}^2$) \\ \hline
Bird model    & 0.97        & 114.84         \\ 
McClear model & 0.99        & 41.04          \\ 
Yang model    & 0.99        & 66.48          \\ \hline
\end{tabular}
\caption{Evaluation of various clear sky models for the sample day in Singapore}
\label{tab:oneday-results}
\end{table}

For a larger statistical duration, we compute such objective measures for all the clear-sky days in the year of 2016. Table~\ref{tab:avg-results} summarizes the results, and shows the average correlation- and RMSE- values. 

\begin{table}[htb]
\centering
\begin{tabular}{l|c|c}
\hline
              & Correlation & RMSE (W/$\mbox{m}^2$) \\ \hline
Bird model    & 0.93       & 152.96         \\ 
McClear model & 0.95        & 106.43         \\ 
Yang model    & 0.95        & 117.98         \\ \hline
\end{tabular}
\caption{Average correlation and RMSE values for all clear-sky days in the year of 2016.}
\label{tab:avg-results}
\end{table}

Table~\ref{tab:avg-results} clearly concludes that the McClear model performs the best across other benchmarking clear-sky radiation models. 

\psection{Conclusion}
In this paper, we have provided a comparative analysis of various state-of-the-art clear sky models. We recorded our observations for the tropical country of Singapore. We recorded all the clear sky days in a single year, and estimated the best model amongst all benchmarking methods. We conclude that the CAMS McClear model is the most accurate clear-sky model for Singapore. In our future work, we plan to integrate images from whole-sky imager for reliable solar energy estimation and forecasting. We also plan to design hardware-efficient sky cameras with low-power for continuous usage, as it is ubiquitous in the areas of biomedical systems~\cite{Deepu2016}. 

\ack
The authors would like to thank Luke Witmer for his python implementation of Bird Clear Sky Model for Solar Radiation Modeling~\cite{Witmer2015}.

\end{paper}

\end{document}